\documentclass[a4paper,UKenglish,cleveref, autoref, thm-restate]{lipics-v2021}

\usepackage{tikz}
\usetikzlibrary{plotmarks}

\title{Simple in-place yet comparison-optimal Mergesort}

\author{Christian Siebert}{University of Stuttgart, High-Performance Computing Center Stuttgart, Germany}{christian.siebert@hlrs.de}{https://orcid.org/0009-0009-8382-7487}{}

\authorrunning{C. Siebert}

\Copyright{Christian Siebert}



\category{}

\relatedversion{}

\nolinenumbers

\EventLongTitle{This publication was submitted to ESA 2024, to SOSA '25 and again to ESA 2025, but got rejected all the time}
\EventShortTitle{Siebert 2025}
\EventAcronym{Siebert}
\EventYear{2025}
\EventDate{December 24--27, 2016}
\EventLocation{Little Whinging, United Kingdom}
\EventLogo{}
\SeriesVolume{42}
\ArticleNo{23}

\begin{document}

\maketitle

\begin{abstract}
   
\emph{Mergesort} is one of the few efficient sorting algorithms and,
despite being the oldest one, often still the method of choice today.
In contrast to some alternative algorithms, it always runs efficiently
using $\mathcal{O}(n \log n)$ element comparisons and usually works in
a stable manner. Its only practical disadvantage is the need for a
second array, and thus twice the amount of memory.
This can be an impeding factor, especially when handling large amounts of data,
where it is often either impractical or even impossible to fall back
to slower or unstable sorting alternatives.
Therefore, many attempts have been made to fix
this problem by adapting \emph{Mergesort} to work \emph{in place}.
While it is known that such algorithms exist, the previously
published solutions are mostly not efficient,
become unstable, and/or are very complex. This renders them
practically useless for real-world applications.

In this paper, we propose a novel \emph{in-place} \emph{Mergesort} algorithm
that is stable by design. Albeit its running time of $\mathcal{O}(n \log^2 n)$
is not quite optimal, it still works efficiently, both in theory using
the optimal number of $\mathcal{O}(n \log n)$ comparisons and in practice
with low constants, while being easily comprehensible.
The baseline for this new algorithm includes just two prerequisites:
1) an optimal array rotation; and 2) the \emph{co-ranking} idea,
published in 2014 and originally intended to parallelize \emph{Mergesort}.
Although it would certainly be possible to parallelize the presented
algorithm, this paper focuses on the sequential aspect of this
efficient, stable and \emph{in-place} \emph{Mergesort} algorithm.
Additionally, we implemented our algorithm and present performance
results measured on one of the largest shared memory systems currently
available.

\end{abstract}

\section{Introduction}
\label{sec:intro}

Sorting is one of the most fundamental problems in computer science
and is used in many applications. A sorting algorithm brings a given set of
elements into a specified order. W.l.o.g., we are concentrating
on data structures that allow random access to their elements, such as
simple arrays. In the following, such arrays are depicted, for example,
as $A[0 \dots n-1]$, where $A$ is the name of the array containing all
elements, and $0$ as well as $n-1$ are the lower and upper bounds---both
inclusive---of element entries, meaning the number of elements in this
array is $n = |A|$. Although there are very specialized sorting algorithms
restricting its use, for example, to small element types (e.g.,
\emph{radix sort}), we consider the generic and unrestricted case based on
comparisons. In such comparison-based sorting algorithms the order of
elements is specified
using a comparison operator, which takes two elements as input, compares
their respective keys, and returns whether the first element either
precedes $\prec$ or succeeds $\succ$ the second element, or if both
have identical keys $=$, i.e., they are equal. For example, in the
standard C library, this comparison operator is a function that uses
a return value of type integer to encode these three options. As such,
a valid comparison function can return -1, 1, and 0, respectively.
This would also define the sorting order to be ascending. On the other
hand, a descending order can be achieved by instead returning
1, -1, and 0, respectively.
Two further terms that are necessary for the
basic understanding of this paper are
stability and rank. A sorting algorithm is said to be stable if
elements with equal keys stay in the same relative order during sorting.
The rank of an element corresponds to the unique index of this element if
all elements have been sorted in a stable manner.

\definecolor{goodcolor}{rgb}{0.1,0.7,0.1}
\definecolor{badcolor}{rgb}{0.9,0.1,0.1}

\newcommand{\good}[1]{#1 \ \includegraphics[width=3mm]{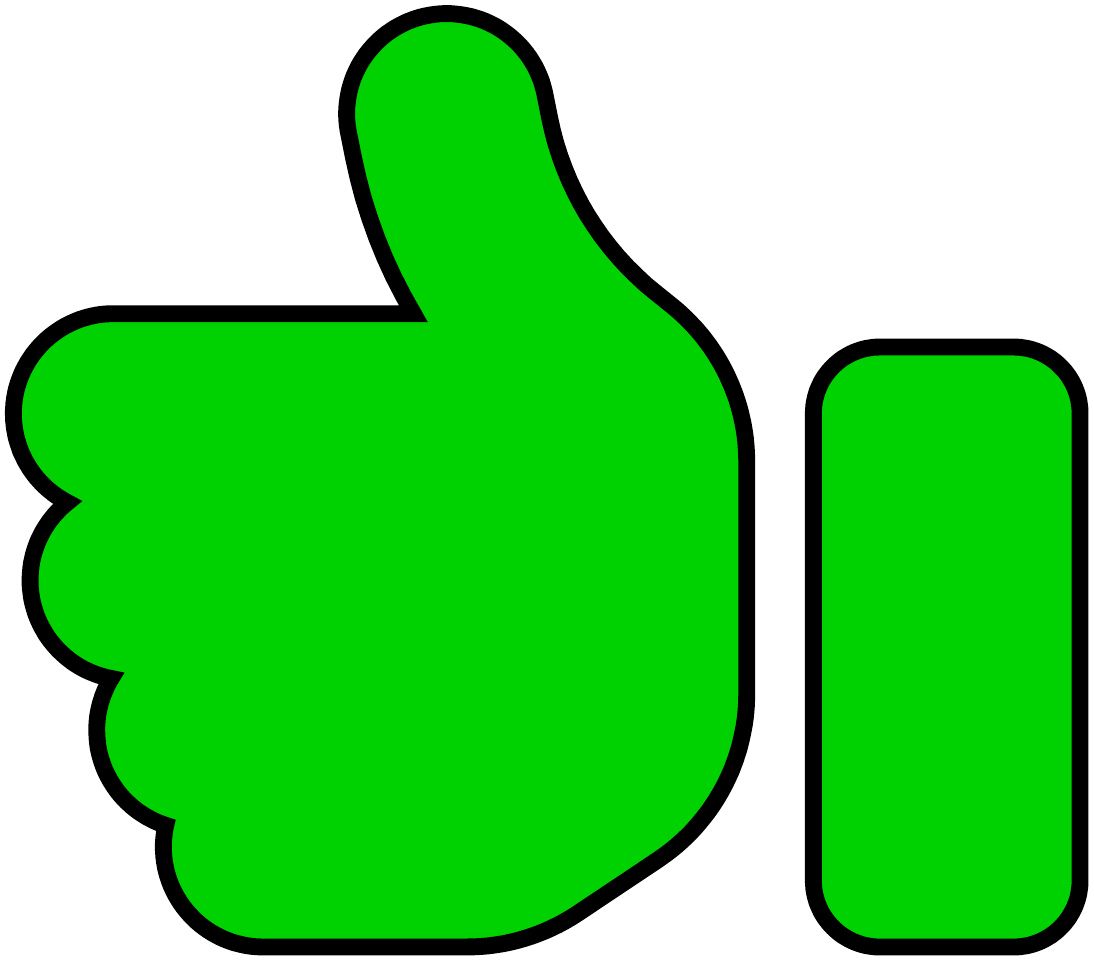} }
\newcommand{\bad}[1]{#1 \ \includegraphics[width=3mm]{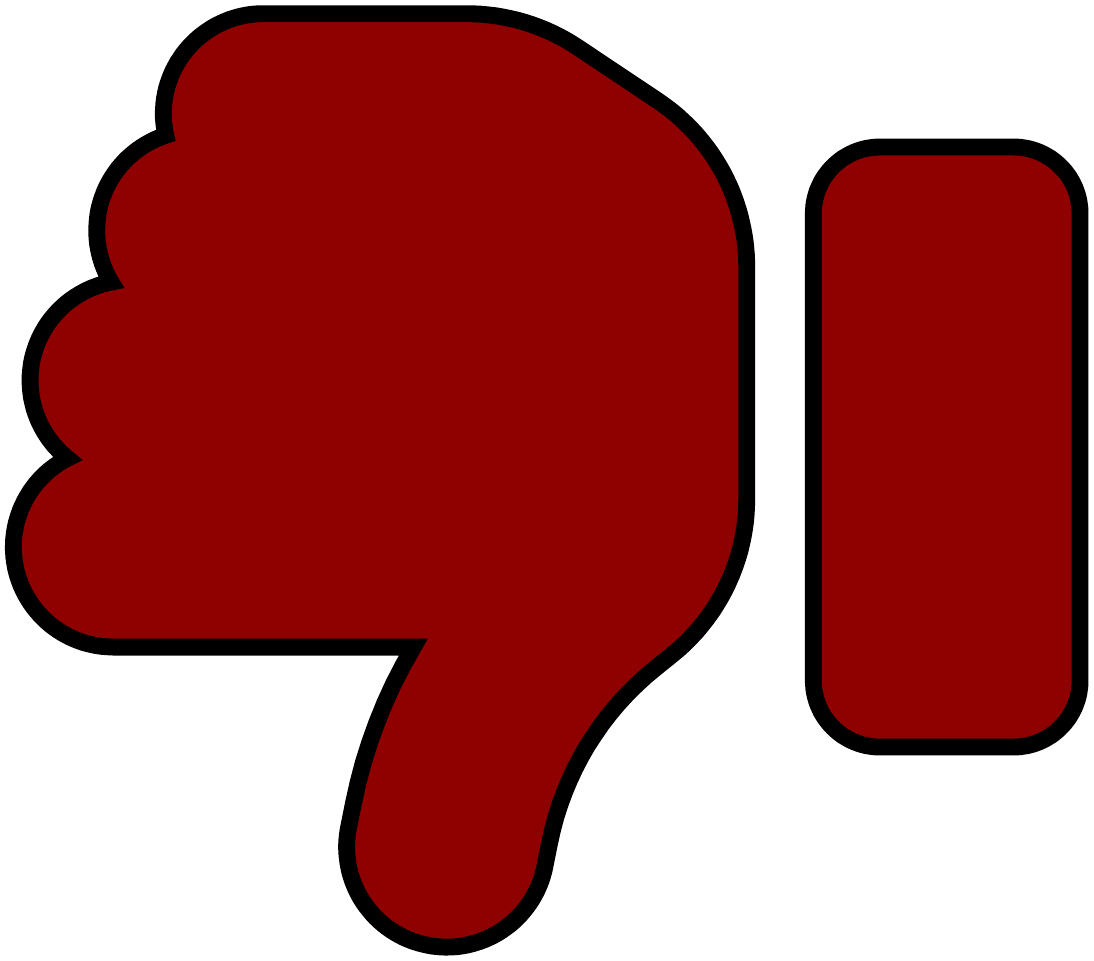} }

It is well known that any comparison-based sorting algorithm needs
$\Omega(n \log n)$ comparisons in the worst case (see, e.g.,
\cite{mybook:intro2algs}, p. 193). As such, we are,
in particular when handling large amounts of data, mostly interested
in sorting algorithms that usually achieve this
theoretical lower bound. While there are
many different sorting algorithms, only a few of them
are appropriate in this context. For instance, the corresponding
Wikipedia page~\cite{wiki:sorting_algorithms} currently lists, explains
and compares more than 45 different sorting algorithms, whereas the
interesting ones for large $n$ 
are represented by \emph{Quicksort},
\emph{Mergesort} and \emph{Heapsort}. These three can also be varied
and combined.
Table~\ref{table:sortprop} compares the aforementioned major
algorithms with respect to their properties and features,
marking these either with a \good{thumbs-up} if this is considered good or
with a \bad{thumbs-down} if not.
In addition to the theoretical performance, we compare also the practical
constant factor, which is ignored in the theoretical $\mathcal{O}$-notation.
This constant can be obtained for a specific implementation and architecture
by measuring its running time for different numbers of elements and
fitting the resulting performance data to the formula $c \cdot n \log n$.

\begin{table}
	\begin{center}
   \begin{tabular}{c|cccc}
      Algorithm & $\sharp$ Comparisons & Constant & Stability & Memory \\
      \hline
      \emph{Quicksort} & \bad{$\mathcal{O}(n^2)$} & \good{small} & \bad{unstable} & \good{$\mathcal{O}(\log n)$} \\
      \emph{Mergesort} & \good{$\mathcal{O}(n \log n)$} & \good{small} & \good{stable} & \bad{$\mathcal{O}(n)$} \\
      \emph{Heapsort} & \good{$\mathcal{O}(n \log n)$} & \bad{large} & \bad{unstable} & \good{$\mathcal{O}(1)$} \\
   \end{tabular}
	\end{center}
   \caption{Properties and features of ``fast'' sorting algorithms}
   \label{table:sortprop}
\end{table}

For example, \emph{Quicksort} is popular because it has, in practice, a
small hidden constant in the theoretical $\mathcal{O}$-notation and achieves
$\mathcal{O}(n \log n)$ comparisons in many cases, such as on average whenever
pivots are picked at random. However, if unsuited pivot values are chosen
(e.g., with fixed pivots and pre-sorted inputs),
it can slow down to as low as $\mathcal{O}(n^2)$, which is catastrophic for
large $n$. \emph{Heapsort},
on the other hand, always achieves $\mathcal{O}(n \log n)$ comparisons
and needs no additional memory but
is cache-inefficient and thus results in larger practical constants.
Additionally, both of these sorting algorithms do not preserve the order of
equal elements, so they are not stable. Although \emph{Mergesort} meets this
stability criterion and is efficient, it works wastefully
regarding additional memory.
In essence, an application programmer can only pick up to three good features
out of the four compared properties and hence has to decide which
sorting algorithm is best suited for a specific application.

While there are attempts to improve upon certain disadvantages,
such as improved pivot-selection strategies for \emph{Quicksort} or
reintroducing stability, this usually worsens one or more
of the other properties. For example, every sorting algorithm can be
turned into a stable sort by adding the original index to each element
and comparing also the indices for equal elements. However, this will
directly lead to an additional memory consumption of $\mathcal{O}(n)$.
This paper concentrates on making
\emph{Mergesort} memory efficient, which is described by the
algorithmic term \emph{in-place}. A formal definition of this term
will be given in Definition~\ref{def:inplace}. The goal is to
create an \emph{in-place} \emph{Mergesort} that is still stable,
works with an optimal number of comparisons and with low constants.
Ideally, it should nevertheless be simple enough to be easily comprehensible.

\subsection{Related work}

There are numerous publications on how to make \emph{Mergesort}
memory efficient. Naive approaches can reduce
the additional space consumption down to $\mathcal{O}(1)$, but at
the cost of drastically increasing the number of comparisons to
$\mathcal{O}(n^2)$ or even $\mathcal{O}(n^2 \log n)$.
Intermediate practical solutions, such as some versions of the
C++ STL library, try to keep the performance as good as possible but
only reduce the additional space consumption to $\mathcal{O}(\sqrt{n})$.
Usually, the complexity
increases whenever space consumption is lowered. There are,
for example, several $\mathcal{O}(n \log^2 n)$ approaches that
can remove the additional space requirement of \emph{Mergesort},
but most of them also lose stability and
are often only of theoretical significance.
Katajainen et al. managed to create an $\mathcal{O}(n \log n)$
\emph{Mergesort} with no additional space
requirements~\cite{paper:Katajainen}. However, their algorithm
is complex, with large hidden constants (p. 9: ``unacceptably slow'',
albeit they improved it later with lots of tuning)
and no stability. Moreover, they raised the significant question of
whether it would be possible to develop a powerful, stable,
\emph{in-place Mergesort}. A whole decade later, in 2006, this question
was still open, and Chen~\cite{paper:Chen} observed:
``Although many stable in-place merges have been published, none has
been shown to be practical.'' 
In 2007, Franceschini settled this long-standing open question and 
introduced an optimal sorting algorithm that works \emph{in place}
while being stable~\cite{paper:opt_stableinplace}. However, this
paper contains 27 pages of purely theoretical work. 
Another 15 years later, in a recent and comprehensive paper,
Axtmann et al. compare various in-place sorting
algorithms~\cite{paper:Axtmann_inplace}, covering many algorithms, features,
implementation/optimization details, and even performance evaluations.
Regarding \emph{in-place Mergesort} algorithms, they provide
only little detail within these $75$ pages, on less than half a percent of the paper space,
and state:
``There is a considerable amount of theory work on in-place sorting.
However, there are few---mostly negative---results of transferring
the theory work into practice.'' Regarding stable versions, they
refer only to WikiSort and GrailSort, and point out
that non-in-place Mergesort is considerably faster than in-place
Mergesort.
WikiSort is based on an in-place and also stable merging
algorithm~\cite{paper:RBmerging} for two input arrays of sizes
$n_A$ and $n_B$, which is furthermore optimal regarding
comparisons and moves. Nevertheless, it only does so for skewed
ratios $k = \frac{n_A}{n_B}$ of $k \geq \sqrt{n_B}$, but the
general \emph{Mergesort} (cf. Algorithm~\ref{alg:Mergesort}) uses
a ratio of $k \approx 1$. Any such skewed ratio will degrade the
overall performance, making it non-optimal. Furthermore, the
implementation uses a default cache size of $512$ elements. Contrary to
the author's claim that ``it's still $\mathcal{O}(1)$ memory'', we argue
that this is more like $\mathcal{O}(\log^2 n)$ for our largest possible
data set, or at least $\mathcal{O}(\log n)$ for $n$  representing all
atoms in the universe, thus not constant. In addition, this algorithm
exploits the presence of $\mathcal{O}(\sqrt{n})$ duplicate data values,
which is a strong restriction and is compelled by the implementation by
sorting only small integer values that are generated by the weak pseudo-random number
generator \texttt{rand()}, which often returns only values between $0$
and $32767$. Thus, just by generating one billion elements or more, the
presence of so many duplicates can be guaranteed. As in practice this
will rarely be the case, this makes it a very specialized algorithm.
Besides, it is well known
that small integers can be sorted much faster, even in linear time, using
other sorting algorithms such as \emph{counting sort} or \emph{radix sort}.
Another weakness of \texttt{rand()} is its small period, which creates
further duplicates when generating larger amounts of data.
Therefore, we opted to handle floating-point numbers and a
better pseudo-random number generator with a much larger period
in our evaluation, which will only rarely
produce duplicate values and therefore does not impose any
such restrictions on the input data.
GrailSort is based on ideas from Huang and
Langston~\cite{paper:stable_merging}. These are not only complex but
also so slow that GrailSort essentially
falls back to buffering $\sqrt{n}$ elements. The implementation
suffers from similar problems and restrictions as WikiSort, namely
a non-constant cache size, handling of only integers, and a
self-written pseudo-random number generator with a small period.
Furthermore, the implementation is header-only, and, therefore,
everything needs to be known at compile time. This makes it inflexible
as a general-purpose library but allows many compiler optimizations.
It also handles cases with small number of elements by a different
sorting algorithm.
Despite these optimizations and even when keeping the restrictions on the
input data, just lowering the cache size results in a much slower
implementation than what we will present in this paper.

This paper proposes a \emph{Mergesort} that works
\emph{in place} while keeping all the good features, namely
$\mathcal{O}(n \log n)$ comparisons with small hidden constants, and
still being stable as well as comprehensible.
As such, we conclude the open question from Katajainen et al.
positively and demonstrate our \emph{Mergesort} to be practical in
contrast to Chen's and Axtmann's observations.

\section{Prerequisites}
\label{sec:prerequisites}

\lstset{mathescape=true,morekeywords={int,while,typeof,if,endwhile,endif,else,typesize,type,for,end,endfor,repeat,until,bool}, morecomment=[l]{//}}

Our novel \emph{in-place Mergesort} will be presented in
Section~\ref{sec:algorithm}. This variant
utilizes just two prerequisites that are introduced in this section.
These already exist and can thus simply be used as ``black box''
functionality. For convenience, we provide pseudocodes
for both of them. As such, it is not necessary for the understanding of
the actual algorithm to also fully comprehend how these two components
operate. Nonetheless, the more interested
reader is encouraged to additionally
explore their respective details as well as the cited publications
given in the next subsections. Naturally, future and practical
improvements or optimizations in any of these prerequisites will directly
lead to a better \emph{in-place Mergesort}.

\subsection{Optimal array rotation}
\label{sec:array_rotation}

First, we need a function to rotate a given array with $n$ elements by
an arbitrary offset. A rotation in this context means that elements that
fall out on one side of the array are placed back into the other
side of the array.
Since the contemplated \emph{Mergesort} algorithm should work
\emph{in place}, this array rotation consequently needs to work
\emph{in place} as well, ideally with $\mathcal{O}(1)$ additional memory.
Moreover, this rotation should also remain time-optimal, meaning it
should run in $\mathcal{O}(n)$ time, which is apparently also the lower bound.

\subsubsection{Insufficient approaches}

If we allowed a temporary array to keep a copy of all elements, such a
rotation would become trivial and run in optimal time $\mathcal{O}(n)$.
This temporary array would then require an additional space
of $\mathcal{O}(n)$, which would clearly not be \emph{in place}, as desired.

On the other hand, if we were to rotate all elements only by one position and
repeat this $r$ times, where $r$ is the rotation offset, it could be
done \emph{in place} but would need $\mathcal{O}(r)$ repetitions. Since
this arbitrary offset can always be mapped between $0 \leq r < n$ using
$r \ \mathrm{mod} \ n$, this would lead to $\mathcal{O}(n)$ rotations
and, overall, to an inefficient $\mathcal{O}(n^2)$ running time.

\subsubsection{Time- and space-optimal rotation}

Instead of moving all array elements only by one position, we could
move them directly by $r$ positions. More formally, starting at
position $s$, this would move all elements at positions
\begin{equation}
	s + k \cdot r \ (\mathrm{mod} \ n) \quad \forall k \in \mathbb{Z}.
\end{equation}

{\bf Case 1:}
Whenever $r$ and $n$ are coprime, i.e., their greatest common divisor
(short $gcd$) is $1$, this would rotate all elements in a single cycle
with a cycle length of $n$.
{\bf Example:} If a given array contains $12$ elements that should be
rotated by an offset of $5$, then there is just a single cycle, written
in cycle notation as $(0 \ 5 \ 10 \ 3 \ 8 \ 1 \ 6 \ 11 \ 4 \ 9 \ 2 \ 7)$.

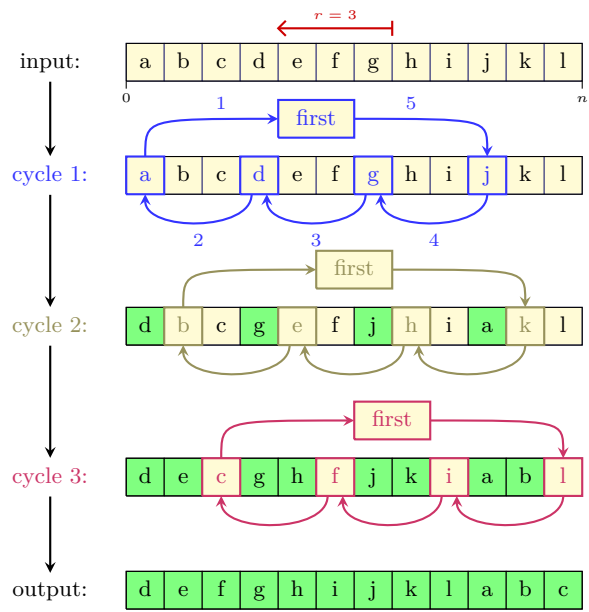
\begin{figure}[htbp]
   \begin{center}
      \begin{tikzpicture}[scale=1.0]
   \definecolor{hlrsblue}{HTML}{BBD037}
   \definecolor{hlrsyelloworg}{HTML}{F8EA42}
   \colorlet{hlrsyellow}{hlrsyelloworg!70!black}
   \definecolor{hlrspurple}{HTML}{01B2E9}
   \definecolor{hlrsred}{HTML}{E84615}

   \colorlet{colbars}{blue!50!white}
   \colorlet{colbarsoff}{gray!50}
   \colorlet{colcloud}{yellow!20!white}
   \colorlet{colranks}{red!80!black}
   \colorlet{colsmall}{blue!50!white}
   \colorlet{collarge}{green!50!white}
   \colorlet{colsmallC}{colsmall!30!white}
   \colorlet{collargeC}{collarge!30!white}

   \colorlet{colRotOffset}{colranks}
   \colorlet{colCompleted}{collarge}
   \colorlet{colCycle1}{blue!80!white}
   \colorlet{colCycle2}{yellow!50!black}
   \colorlet{colCycle3}{purple!80!white}

   \foreach \x/\val in {1/a,2/b,3/c,4/d,5/e,6/f,7/g,8/h,9/i,10/j,11/k,12/l}
   {
      \draw[colsmall!50!black,fill=colcloud] (\x cm*0.5,8.5cm) --
         (0.5cm + \x cm*0.5,8.5cm) -- (0.5cm + \x cm*0.5,9.0cm) --
         (\x cm*0.5,9.0cm) -- (\x cm*0.5,8.5cm);
      \node[font=\footnotesize,color=black,anchor=mid] at (0.25cm + \x cm*0.5,8.75cm) {\val};
   }
   \draw[black,fill=none] (0.5cm,8.5cm) --
         (6.5cm,8.5cm) -- (6.5cm,9.0cm) --
         (0.5cm,9.0cm) -- (0.5cm,8.5cm);
   \node[font=\footnotesize,color=black,anchor=mid] (input) at (-0.5cm,8.75cm) {input:};
   \node[font=\tiny,color=black,anchor=mid] at (0.5cm,8.3cm) {$0$};
   \draw[black,fill=none] (0.5cm,8.5cm) -- (0.5cm,8.4cm);
   \node[font=\tiny,color=black,anchor=mid] at (6.5cm,8.3cm) {$n$};
   \draw[black,fill=none] (6.5cm,8.5cm) -- (6.5cm,8.4cm);
   \draw[colRotOffset,thick] (2.5cm,9.2cm) -- (4.0cm,9.2cm);
   \draw[colRotOffset,thick] (4.0cm,9.1cm) -- (4.0cm,9.3cm);
   \draw[colRotOffset,thick] (2.6cm,9.1cm) -- (2.5cm,9.2cm) -- (2.6cm,9.3cm);
   \node[font=\tiny,color=colRotOffset,anchor=mid] at (3.25cm,9.4cm) {$r = 3$};

   \foreach \x/\val in {2/b,3/c,5/e,6/f,8/h,9/i,11/k,12/l}
   {
      \draw[colsmall!50!black,fill=colcloud] (\x cm*0.5,7.0cm) --
         (0.5cm + \x cm*0.5,7.0cm) -- (0.5cm + \x cm*0.5,7.5cm) --
         (\x cm*0.5,7.5cm) -- (\x cm*0.5,7.0cm);
      \node[font=\footnotesize,color=black,anchor=mid] at (0.25cm + \x cm*0.5,7.25cm) {\val};
   }
   \draw[black,fill=none] (0.5cm,7.0cm) --
         (6.5cm,7.0cm) -- (6.5cm,7.5cm) --
         (0.5cm,7.5cm) -- (0.5cm,7.0cm);
   \draw[colCycle1,fill=colcloud,thick] (2.5cm,7.75cm) --
         (3.5cm,7.75cm) -- (3.5cm,8.25cm) --
         (2.5cm,8.25cm) -- (2.5cm,7.75cm);
   \node[font=\footnotesize,color=colCycle1,anchor=mid] at (3.0cm,8.0cm) {first};
   \foreach \x/\val in {1/a,4/d,7/g,10/j}
   {
      \draw[colCycle1,fill=colcloud,thick] (\x cm*0.5,7.0cm) --
         (0.5cm + \x cm*0.5,7.0cm) -- (0.5cm + \x cm*0.5,7.5cm) --
         (\x cm*0.5,7.5cm) -- (\x cm*0.5,7.0cm);
      \node[font=\footnotesize,color=colCycle1,anchor=base] at (0.25cm + \x cm*0.5,7.175cm) {\val};
   }
   \draw[-stealth,colCycle1,thick] (0.75cm,7.5cm) .. controls +(up:0.5cm) and +(left:1.75cm) .. (2.5cm,8.0cm);
   \draw[-stealth,colCycle1,thick] (3.5cm,8.0cm) .. controls +(right:1.75cm) and +(up:0.5cm) .. (5.25cm,7.5cm);
   \draw[-stealth,colCycle1,thick] (5.25cm,7.0cm) .. controls +(down:0.5cm) and +(down:0.5cm) .. (3.85cm,7.0cm);
   \draw[-stealth,colCycle1,thick] (3.65cm,7.0cm) .. controls +(down:0.5cm) and +(down:0.5cm) .. (2.35cm,7.0cm);
   \draw[-stealth,colCycle1,thick] (2.15cm,7.0cm) .. controls +(down:0.5cm) and +(down:0.5cm) .. (0.75cm,7.0cm);
   \node[font=\footnotesize,color=colCycle1,anchor=mid] (cycle1) at (-0.5cm,7.25cm) {cycle $1$:};
   \node[font=\scriptsize,color=colCycle1,anchor=mid] at (1.75cm,8.2cm) {$1$};
   \node[font=\scriptsize,color=colCycle1,anchor=north] at (1.45cm,6.6cm) {$2$};
   \node[font=\scriptsize,color=colCycle1,anchor=north] at (3.00cm,6.6cm) {$3$};
   \node[font=\scriptsize,color=colCycle1,anchor=north] at (4.55cm,6.6cm) {$4$};
   \node[font=\scriptsize,color=colCycle1,anchor=mid] at (4.25cm,8.2cm) {$5$};

   \foreach \x/\val in {3/c,6/f,9/i,12/l}
   {
      \draw[colsmall!50!black,fill=colcloud] (\x cm*0.5,5.0cm) --
         (0.5cm + \x cm*0.5,5.0cm) -- (0.5cm + \x cm*0.5,5.5cm) --
         (\x cm*0.5,5.5cm) -- (\x cm*0.5,5.0cm);
      \node[font=\footnotesize,color=black,anchor=mid] at (0.25cm + \x cm*0.5,5.25cm) {\val};
   }
   \foreach \x/\val in {1/d,4/g,7/j,10/a}
   {
      \draw[black,fill=colCompleted] (\x cm*0.5,5.0cm) --
         (0.5cm + \x cm*0.5,5.0cm) -- (0.5cm + \x cm*0.5,5.5cm) --
         (\x cm*0.5,5.5cm) -- (\x cm*0.5,5.0cm);
      \node[font=\footnotesize,color=black,anchor=mid] at (0.25cm + \x cm*0.5,5.25cm) {\val};

   }
   \draw[black,fill=none] (0.5cm,5.0cm) --
         (6.5cm,5.0cm) -- (6.5cm,5.5cm) --
         (0.5cm,5.5cm) -- (0.5cm,5.0cm);
   \draw[colCycle2,fill=colcloud,thick] (3.0cm,5.75cm) --
         (4.0cm,5.75cm) -- (4.0cm,6.25cm) --
         (3.0cm,6.25cm) -- (3.0cm,5.75cm);
   \node[font=\footnotesize,color=colCycle2,anchor=mid] at (3.5cm,6.0cm) {first};
   \foreach \x/\val in {2/b,5/e,8/h,11/k}
   {
      \draw[colCycle2,fill=colcloud,thick] (\x cm*0.5,5.0cm) --
         (0.5cm + \x cm*0.5,5.0cm) -- (0.5cm + \x cm*0.5,5.5cm) --
         (\x cm*0.5,5.5cm) -- (\x cm*0.5,5.0cm);
      \node[font=\footnotesize,color=colCycle2,anchor=base] at (0.25cm + \x cm*0.5,5.175cm) {\val};
   }
   \draw[-stealth,colCycle2,thick] (1.25cm,5.5cm) .. controls +(up:0.5cm) and +(left:1.75cm) .. (3.0cm,6.0cm);
   \draw[-stealth,colCycle2,thick] (4.0cm,6.0cm) .. controls +(right:1.75cm) and +(up:0.5cm) .. (5.75cm,5.5cm);
   \draw[-stealth,colCycle2,thick] (5.75cm,5.0cm) .. controls +(down:0.5cm) and +(down:0.5cm) .. (4.35cm,5.0cm);
   \draw[-stealth,colCycle2,thick] (4.15cm,5.0cm) .. controls +(down:0.5cm) and +(down:0.5cm) .. (2.85cm,5.0cm);
   \draw[-stealth,colCycle2,thick] (2.65cm,5.0cm) .. controls +(down:0.5cm) and +(down:0.5cm) .. (1.25cm,5.0cm);
   \node[font=\footnotesize,color=colCycle2,anchor=mid] (cycle2) at (-0.5cm,5.25cm) {cycle $2$:};

   \foreach \x/\val in {1/d,2/e,4/g,5/h,7/j,8/k,10/a,11/b}
   {
      \draw[black,fill=colCompleted] (\x cm*0.5,3.0cm) --
         (0.5cm + \x cm*0.5,3.0cm) -- (0.5cm + \x cm*0.5,3.5cm) --
         (\x cm*0.5,3.5cm) -- (\x cm*0.5,3.0cm);
      \node[font=\footnotesize,color=black,anchor=mid] at (0.25cm + \x cm*0.5,3.25cm) {\val};
   }
   \draw[black,fill=none] (0.5cm,3.0cm) --
         (6.5cm,3.0cm) -- (6.5cm,3.5cm) --
         (0.5cm,3.5cm) -- (0.5cm,3.0cm);
   \draw[colCycle3,fill=colcloud,thick] (3.5cm,3.75cm) --
         (4.5cm,3.75cm) -- (4.5cm,4.25cm) --
         (3.5cm,4.25cm) -- (3.5cm,3.75cm);
   \node[font=\footnotesize,color=colCycle3,anchor=mid] at (4.0cm,4.0cm) {first};
   \foreach \x/\val in {3/c,6/f,9/i,12/l}
   {
      \draw[colCycle3,fill=colcloud,thick] (\x cm*0.5,3.0cm) --
         (0.5cm + \x cm*0.5,3.0cm) -- (0.5cm + \x cm*0.5,3.5cm) --
         (\x cm*0.5,3.5cm) -- (\x cm*0.5,3.0cm);
      \node[font=\footnotesize,color=colCycle3,anchor=base] at (0.25cm + \x cm*0.5,3.175cm) {\val};
   }
   \draw[-stealth,colCycle3,thick] (1.75cm,3.5cm) .. controls +(up:0.5cm) and +(left:1.75cm) .. (3.5cm,4.0cm);
   \draw[-stealth,colCycle3,thick] (4.5cm,4.0cm) .. controls +(right:1.75cm) and +(up:0.5cm) .. (6.25cm,3.5cm);
   \draw[-stealth,colCycle3,thick] (6.25cm,3.0cm) .. controls +(down:0.5cm) and +(down:0.5cm) .. (4.85cm,3.0cm);
   \draw[-stealth,colCycle3,thick] (4.65cm,3.0cm) .. controls +(down:0.5cm) and +(down:0.5cm) .. (3.35cm,3.0cm);
   \draw[-stealth,colCycle3,thick] (3.15cm,3.0cm) .. controls +(down:0.5cm) and +(down:0.5cm) .. (1.75cm,3.0cm);
   \node[font=\footnotesize,color=colCycle3,anchor=mid] (cycle3) at (-0.5cm,3.25cm) {cycle $3$:};

   \foreach \x/\val in {1/d,2/e,3/f,4/g,5/h,6/i,7/j,8/k,9/l,10/a,11/b,12/c}
   {
      \draw[black,fill=colCompleted] (\x cm*0.5,1.5cm) --
         (0.5cm + \x cm*0.5,1.5cm) -- (0.5cm + \x cm*0.5,2.0cm) --
         (\x cm*0.5,2.0cm) -- (\x cm*0.5,1.5cm);
      \node[font=\footnotesize,color=black,anchor=mid] at (0.25cm + \x cm*0.5,1.75cm) {\val};
   }
   \draw[black,fill=none] (0.5cm,1.5cm) --
         (6.5cm,1.5cm) -- (6.5cm,2.0cm) --
         (0.5cm,2.0cm) -- (0.5cm,1.5cm);
   \node[font=\footnotesize,color=black,anchor=mid] (output) at (-0.5cm,1.75cm) {output:};

   \draw[-stealth,black,thick] (input.south) -> (cycle1.north);
   \draw[-stealth,black,thick] (cycle1.south) -> (cycle2.north);
   \draw[-stealth,black,thick] (cycle2.south) -> (cycle3.north);
   \draw[-stealth,black,thick] (cycle3.south) -> (output.north);

\end{tikzpicture}
   \end{center}
   \caption{Optimal algorithm to rotate a given array with $12$ elements by an offset of $3$}
   \label{fig:rotation_howto}
\end{figure}

{\bf Case 2:} Whenever $r$ and $n$ are not
coprime, they have the greatest divisor $d = \gcd(n, r)$ in common.
In that case, there exist $d$ cycles with $n / d$ elements each.
Regardless of where cycles are started, elements from different
cycles are always distinct.
{\bf Example:} If the array contains again $12$ elements but should be
rotated by a different offset of $3$, then there are $\gcd(12, 3) = 3$
cycles with $12 / 3 = 4$ elements each, namely
$(0 \ 3 \ 6 \ 9) (1 \ 4 \ 7 \ 10) (2 \ 5 \ 8 \ 11)$.
Such an example is visualized in Figure~\ref{fig:rotation_howto}.

Since we have $d$ cycles with $n / d$ elements, each and all
elements are different, so we rotate $d \times n / d = n$ elements
in total. Therefore, all elements are rotated correctly in
optimal time and space. This algorithm
was presented by Fletcher and Silver in 1965 and is
often referred to as the ``juggling algorithm''~\cite[p.~209]{mybook:Stepanov}.

There are several other rotation algorithms, such as the
``triple reversal rotation'' and more complex variations of it (e.g.,
Gries-Mills rotation, the Grail rotation, the Helix rotation or
the Trinity rotation), which improve locality but at the cost
of touching every element usually three times instead of only once.
Although this sounds bad in theory, it can be beneficial in terms of
practical performance on some architectures or certain input distributions.
The presented rotation algorithm was chosen because it is
not only simpler but also gives recurring good enough performance
regardless of the architecture or given data distribution.

\subsubsection{Optimizations}

Although useful for a simplified understanding, we do not need to calculate
this greatest common divisor for an efficient implementation. It should be
pointed out that any cycle always ends at its starting position $s$.
Thus, we can simply increase the starting position whenever we
detect such a cycle's end and proceed to the next cycle. We also do not
need to calculate the number of cycles in advance because we know that
all array elements are moved exactly once, either directly or indirectly
through a single temporary variable (here \texttt{first}), to their final destination.
A simple loop over all elements, only with an adjusted starting
index whenever a cycle end is detected, is sufficient, as depicted in
the pseudocode given in Algorithm~\ref{alg:code_rotate}.
These optimizations slightly enhance the STL rotation algorithms
given in~\cite{paper:rotate}.

\begin{note}\label{note:rotate_leftright}
   A rotation in the opposite direction can be achieved with a
   negative offset $r$, which can also be done with the corresponding
   positive offset $n - r$. For simplicity reasons, the given pseudocode
   works for $0 \leq r < n$ and rotates the given array to the left.
\end{note}

\begin{lstlisting}[basicstyle=\footnotesize,caption={optimal\_rotate$(A,r)$ rotates a given array},label={alg:code_rotate},captionpos=t,float,abovecaptionskip=-\medskipamount]
// optimal_rotate$(A,r)$ rotates an array $A[]$ with $n$ elements
// by an arbitrary offset $r$ (with $0 \leq r < n$) to the left.
// It runs in optimal time $\mathcal{O}(n)$ using only $\mathcal{O}(1)$ additional space.
int $\mbox{work}, s, i, n, \mbox{next}$
type $\mbox{first}$

$n \gets |A|$ // number of elements in $A$
if $(n > 0)$ && $(r > 0)$
   $\mbox{work} \gets n$ // number of elements to move
   for $s \gets 0; \mbox{work} > 0; s\mathbin{{+}{+}}$
      $i \gets s$
      $\mbox{first} \gets A[s]$ // keep "old" first element of cycle
      repeat
	 $\mbox{next} \gets i + r$
	 if $\mbox{next} \geq n$
	    $\mbox{next} \gets \mbox{next} - n$ // handle wrap-around
	 end if
	 if $\mbox{next} == \mbox{s}$
	    $A[i] \gets \mbox{first}$ // last element in cycle
	 else
	    $A[i] \gets A[\mbox{next}]$ // move one element
	 end if
	 $i \gets \mbox{next}$
	 $\mbox{work}\mathbin{{-}{-}}$
      until $i == s$ // Cycle completed?
   end for
end if
\end{lstlisting}

As this optimal rotation algorithm moves every element exactly once,
its running time is trivially bounded by $\mathcal{O}(n)$. Since it uses
only a few basic variables, it needs $\mathcal{O}(1)$ additional space to
do so.
It is important to note that this array rotation only moves, i.e., copies,
data elements, but it does not do any element comparisons at all.

\subsection{Co-ranking}
\label{sec:co-ranking}

It has been commonly known for at least 30 years how to find the shared
median element of two given sorted arrays in logarithmic time (see, e.g.,
exercise 9.3-8 in~\cite{mybook:intro2algs} on page 223).
However, in the case of \emph{Mergesort}, we do not want to split an
array at the median element but instead at the middle position
(cf. Algorithm~\ref{alg:Mergesort}).
Although somewhat similar in nature, utilizing such a median solution
becomes difficult, especially when also considering duplicate keys
and thus ensuring stability. As such, finding the corresponding
indices of this middle position within the two given arrays seemed
harder, i.e., slower than logarithmic time, before and was sometimes
even disputed as being possible in an efficient way. The relatively new
\emph{co-ranking} algorithm solves this problem directly and will thus
be the second prerequisite that we utilize for our \emph{in-place}
\emph{Mergesort}.
The theoretical foundations of \emph{co-ranking} are published
in~\cite{paper:MergeTheory}, and various
practical implementations for distributed memory machines using
\emph{MPI} one-sided communications are presented in~\cite{paper:MPImerge}.
This algorithm was originally developed to parallelize the \emph{Merge}
step of \emph{Mergesort} efficiently. Only later did we realize
that it can also be used to create an \emph{in-place} version
of \emph{Mergesort}. Although the parallel
\emph{Merge} algorithm and our \emph{in-place Merge} algorithm use
this \emph{co-ranking}, they use it for a completely different purpose.

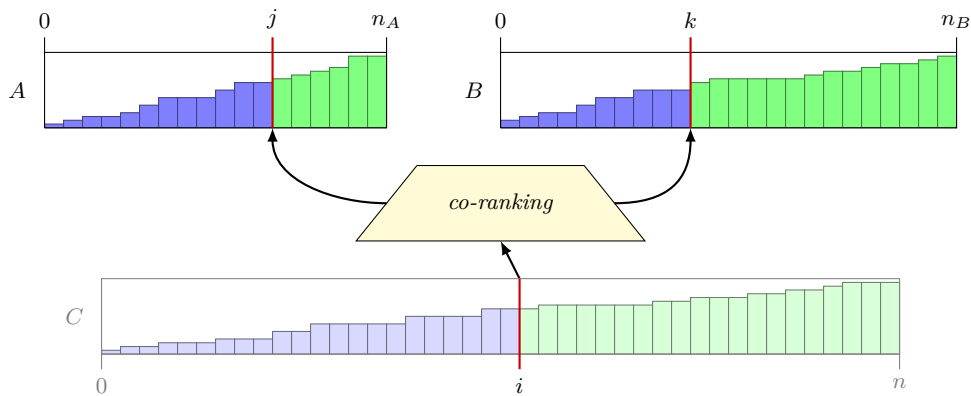
\begin{figure}[htbp]
   \begin{center}
      \begin{tikzpicture}[scale=1.0]
   \colorlet{colbars}{blue!50!white}	
   \colorlet{colbarsoff}{gray!50}
   \colorlet{colcloud}{yellow!20!white}
   \colorlet{colranks}{red!80!black}
   \colorlet{colsmall}{blue!50!white}
   \colorlet{collarge}{green!50!white}
   \colorlet{colsmallC}{colsmall!30!white}
   \colorlet{collargeC}{collarge!30!white}

   \node[font=\footnotesize,color=black,anchor=west] at (0.1cm,0.5cm) {$A$};
   \foreach \x/\val in {
	1/0.05,2/0.10,3/0.15,4/0.15,5/0.20,6/0.30,7/0.40,8/0.40,9/0.40,
	10/0.50,11/0.60,12/0.60}
   {
      \draw[colsmall!50!black,fill=colsmall] (0.45cm + \x cm*0.25,0.0cm) --
		(0.7cm + \x cm*0.25,0.0cm) -- (0.7cm + \x cm*0.25,\val cm) --
		(0.45cm + \x cm*0.25,\val cm) -- (0.45cm + \x cm*0.25,0.0cm);
   }
   \foreach \x/\val in {13/0.65,14/0.70,15/0.75,16/0.80,17/0.95,18/0.95}
   {
      \draw[collarge!50!black,fill=collarge] (0.45cm + \x cm*0.25,0.0cm) --
		(0.7cm + \x cm*0.25,0.0cm) -- (0.7cm + \x cm*0.25,\val cm) --
		(0.45cm + \x cm*0.25,\val cm) -- (0.45cm + \x cm*0.25,0.0cm);
   }
   \draw[black,fill=none] (0.7cm,0.0cm) -- (5.2cm,0.0cm) -- (5.2cm,1.0cm) --
			   (0.7cm,1.0cm) -- (0.7cm,0.0cm);
   \draw[black] (0.7cm,1.2cm) -- (0.7cm,1.0cm);
   \node[font=\footnotesize,color=black,anchor=south] at (0.7cm,1.2cm) {$0$};
   \draw[black] (5.2cm,1.2cm) -- (5.2cm,1.0cm);
   \node[font=\footnotesize,color=black,anchor=south] at (5.2cm,1.2cm) {$n_A$};
   \draw[colranks,thick] (0.7cm + 12cm*0.25,1.2cm) -- (0.7cm + 12cm*0.25,0.0cm);
   \node[font=\footnotesize,color=black,anchor=south] at (0.7cm + 12cm*0.25,1.2cm) {$j$};
   \node[font=\footnotesize,color=black,anchor=west] at (6.1cm,0.5cm) {$B$};
   \foreach \x/\val in {1/0.10,2/0.15,3/0.20,4/0.20,5/0.30,6/0.40,7/0.40,8/0.50,9/0.50,10/0.50}
   {
      \draw[colsmall!50!black,fill=colsmall] (6.45cm + \x cm*0.25,0.0cm) --
                (6.7cm + \x cm*0.25,0.0cm) -- (6.7cm + \x cm*0.25,\val cm) --
                (6.45cm + \x cm*0.25,\val cm) -- (6.45cm + \x cm*0.25,0.0cm);
   }
   \foreach \x/\val in {
	11/0.60,12/0.65,13/0.65,14/0.65,15/0.65,16/0.65,17/0.70,
	18/0.75,19/0.75,20/0.80,21/0.85,22/0.85,23/0.90,24/0.95}
   {
      \draw[collarge!50!black,fill=collarge] (6.45cm + \x cm*0.25,0.0cm) --
                (6.7cm + \x cm*0.25,0.0cm) -- (6.7cm + \x cm*0.25,\val cm) --
                (6.45cm + \x cm*0.25,\val cm) -- (6.45cm + \x cm*0.25,0.0cm);
   }
   \draw[black,fill=none] (6.7cm,0.0cm) -- (12.7cm,0.0cm) -- (12.7cm,1.0cm) --
			   (6.7cm,1.0cm) -- (6.7cm,0.0cm);
   \draw[black] (6.7cm,1.2cm) -- (6.7cm,1.0cm);
   \node[font=\footnotesize,color=black,anchor=south] at (6.7cm,1.2cm) {$0$};
   \draw[black] (12.7cm,1.2cm) -- (12.7cm,1.0cm);
   \node[font=\footnotesize,color=black,anchor=south] at (12.7cm,1.2cm) {$n_B$};
   \draw[colranks,thick] (6.7cm + 10cm*0.25,1.2cm) -- (6.7cm + 10cm*0.25,0.0cm);
   \node[font=\footnotesize,color=black,anchor=south] at (6.7cm + 10cm*0.25,1.2cm) {$k$};

   \draw[fill=colcloud] (5.6cm,-0.5cm) -- (4.8cm,-1.5cm) -- (8.6cm,-1.5cm) -- (7.8cm,-0.5cm)
			-- (5.6cm,-0.5cm);
   \node[font=\footnotesize,fill=none] at (6.7cm,-1.0cm) {\emph{co-ranking}};
   \draw[latex-,thick] (0.7cm + 12cm*0.25,0.0cm) .. controls +(0.0cm,-0.7cm) and +(-0.7cm,0.0cm) .. (5.2cm,-1.0cm);
   \draw[latex-,thick] (6.7cm + 10cm*0.25,0.0cm) .. controls +(0.0cm,-0.7cm) and +(0.7cm,0.0cm) .. (8.2cm,-1.0cm);
   \draw[latex-,thick] (6.7cm,-1.5cm) -- (1.45cm + 22cm*0.25,-2.0cm);

   \node[font=\footnotesize,color=gray,anchor=west] at (0.85cm,-2.5cm) {$C$};
   \foreach \x/\val in {
	1/0.05,2/0.10,3/0.10,4/0.15,5/0.15,6/0.15,7/0.20,8/0.20,9/0.20,
	10/0.30,11/0.30,12/0.40,13/0.40,14/0.40,15/0.40,16/0.40,17/0.50,
	18/0.50,19/0.50,20/0.50,21/0.60,22/0.60}
   {
      \draw[colsmallC!50!black,fill=colsmallC] (1.2cm + \x cm*0.25,-3.0cm) --
                (1.45cm + \x cm*0.25,-3.0cm) -- (1.45cm + \x cm*0.25,\val cm -3.0cm) --
                (1.2cm + \x cm*0.25,\val cm -3.0cm) -- (1.2cm + \x cm*0.25,-3.0cm);
   }
   \foreach \x/\val in {
	23/0.60,24/0.65,25/0.65,26/0.65,27/0.65,28/0.65,29/0.65,
	30/0.70,31/0.70,32/0.75,33/0.75,34/0.75,35/0.80,36/0.80,
	37/0.85,38/0.85,39/0.90,40/0.95,41/0.95,42/0.95}
   {
      \draw[collargeC!50!black,fill=collargeC] (1.2cm + \x cm*0.25,-3.0cm) --
                (1.45cm + \x cm*0.25,-3.0cm) -- (1.45cm + \x cm*0.25,\val cm -3.0cm) --
                (1.2cm + \x cm*0.25,\val cm -3.0cm) -- (1.2cm + \x cm*0.25,-3.0cm);
   }
   \draw[gray,fill=none] (1.45cm,-3.0cm) -- (11.95cm,-3.0cm) --
		(11.95cm,-2.0cm) -- (1.45cm,-2.0cm) -- (1.45cm,-3.0cm);
   \draw[gray] (1.45cm,-3.0cm) -- (1.45cm,-3.2cm);
   \node[font=\footnotesize,color=gray,anchor=north] at (1.45cm,-3.2cm) {$0$};
   \draw[gray] (11.95cm,-3.0cm) -- (11.95cm,-3.2cm);
   \node[font=\footnotesize,color=gray,anchor=north] at (11.95cm,-3.2cm) {$n$};
   \draw[colranks,thick] (1.45cm + 22cm*0.25,-2.0cm) -- (1.45cm + 22cm*0.25,-3.2cm);
   \node[font=\footnotesize,color=black,anchor=north] at (1.45cm + 22cm*0.25,-3.2cm) {$i$};

\end{tikzpicture}
   \end{center}
   \caption{Co-ranking idea for inputs $A$, $B$ and $i$ with results $j$ and $k$}
   \label{fig:co-ranking}
\end{figure}

The general idea of \emph{co-ranking} is visualized in
Figure~\ref{fig:co-ranking}. As input, it takes two presorted arrays
$A$ and $B$, as well as an arbitrary rank (index) $i$. We assume
array $C$  to be the result of merging both input arrays. When
$A$ contains $n_A$ elements and $B$ contains $n_B$ elements, this
array $C$ would need to hold all these elements, i.e., $n = n_A + n_B$.
If $i$ is a valid rank in $C$ (i.e., $0 \leq i \leq n$), the
\emph{co-ranking} functionality will determine the corresponding ranks of
$i$ in the input arrays $A$ and $B$, respectively. Astonishingly, these
\emph{co-ranks} $j$ and $k$ are always unique, and they can be determined
much faster (i.e., $\mathcal{O}(\log n)$ vs. $\mathcal{O}(n)$ time)
{\bf without} actually merging the input arrays and hence without the need
for the result array $C$. This is accomplished in
a binary search fashion while maintaining the invariant $j + k = i$.
It converges lower and upper bounds in both input arrays, where
$j_{\rm low} \leq j$ and $k_{\rm low} \leq k$. The algorithm terminates
when both lemma conditions (see Lemma 2.1 in \cite{paper:MergeTheory}) are
fulfilled, and thus the unique co-ranks $j$ and $k$ are found.

Previously published implementations of the \emph{co-ranking}
functionality add some complexity caused by the nature of
distributed memory systems.
However, since we are only interested in a sequential
implementation in this work, a simpler implementation of the
\emph{co-ranking} functionality is possible and presented as
pseudocode\footnote{Contrary to earlier representations of this
algorithm in \cite{paper:MPImerge} and \cite{paper:MergeTheory},
we replace here the {\bf smaller} / {\bf larger than} signs
(i.e., $<$ / $>$) with {\bf precedes} / {\bf succeeds} symbols
(i.e., $\prec$ / $\succ$), respectively, to depict more clearly,
in contrast to mere range checks, where actual element comparisons
happen.} in Algorithm~\ref{alg:coranking}.
The correctness\footnote{Note that the {\bf \&\&} in the {\bf if} /
{\bf else if} statements denotes
a ``conditional and'' as in the {\bf C/C++} language, and means that the
condition on the right is only evaluated if the condition on the left
evaluates to {\bf $\mbox{true}$} (a.k.a. ``short-circuiting operator'').
}
 of Algorithm~\ref{alg:coranking}, the two referred lemma conditions,
and its run-time analysis are already proven and explained in detail
in~\cite{paper:MergeTheory} and thus omitted here for brevity.

\begin{lstlisting}[basicstyle=\footnotesize,caption={Co\_rank($i, A, B$) determines the two co-ranks of $i$},label={alg:coranking},captionpos=t,float,abovecaptionskip=-\medskipamount]
// Co_rank($i, A, B$) determines for any valid rank (index) $i$ in $C$
// the unique co-ranks $j$ and $k$ in sorted arrays $A$ and $B$
// without having to actually merge $A$ and $B$.
// It runs in $\mathcal{O}(\log n)$ time using only $\mathcal{O}(1)$ additional space.
int $n_A, n_B, j, k, j_{\rm low}, k_{\rm low}$
bool $\mbox{found}$

$(n_A,n_B) \gets (|A|,|B|)$ // number of elements in $A$ and $B$
$j \gets \mbox{min}(i,n_A)$
$k \gets i-j$ // invariant: $j+k=i$
$j_{\rm low} \gets \mbox{max}(0,i-n_B)$
// $k_{\rm low}$ gets initialized within the very first iteration
$\mbox{found} \gets \mbox{false}$
repeat
   // converge indices towards the co-ranks
   if $(j>0$ && $k<n_B)$ && $(A[j-1] \succ B[k])$
      // first Lemma condition violated
      $\delta \gets \lceil \frac{j-j_{\rm low}}{2} \rceil$
      $k_{\rm low} \gets k$
      $(j,k) \gets (j-\delta,k+\delta)$ // decrease $j$
   else if $(k>0$ && $j<n_A)$ && $(B[k-1] \succeq A[j])$
      // second Lemma condition violated
      $\delta \gets \lceil \frac{k-k_{\rm low}}{2} \rceil$
      $j_{\rm low} \gets j$
      $(j,k) \gets (j+\delta,k-\delta)$ // decrease $k$
   else
      // no conditions violated
      $\mbox{found} \gets \mbox{true}$ // unique co-ranks $j$ and $k$ found
   end if
until $\mbox{found}$
$\mbox{return} \ (j,k)$
\end{lstlisting}

As a side note to the introducing sentences regarding the commonly known
median selection, it can be revealed that the problem of selecting
even an arbitrary element at position $i$ (for $0 \leq i < n$) for two
sorted arrays can indeed be solved efficiently in logarithmic time. This
can be achieved easily by employing \emph{co-ranking} for this $i$
and then returning either the minimum of $A[j]$ and $B[k]$ if both exist
or the existing element of these. We leave the details as an exercise for
the interested reader and instead continue to utilize the powerful
\emph{co-ranking} algorithm differently to create the
desired \emph{in-place Mergesort}.

\section{In-place Mergesort}
\label{sec:algorithm}

   The traditional top-down implementation of \emph{Mergesort} uses recursion,
as shown in Algorithm~\ref{alg:Mergesort}. The problematic part lies within
its \emph{Merge} routine, drafted in Algorithm~\ref{alg:Merge_buffered},
which typically allocates a temporary buffer of the same size
as the original input array, merges two array halves therein, and copies
the results back into the input array. This strategy obviously requires
$\mathcal{O}(n)$ additional memory and thus does not work \emph{in place}.
It should be noted that the array halves used in \emph{Mergesort} are always
direct neighbors, i.e., the first array ends where the second array starts.
This property will help us later create an \emph{in-place} variant of
\emph{Merge}. For this reason, \emph{Merge}() can either call
\emph{Merge}\_buffered($A, mid, n-mid$) or
\emph{Merge}\_inplace($A, mid, n-mid$). Consequently, an
\emph{in-place Merge} will directly lead to an \emph{in-place Mergesort}.

\begin{lstlisting}[basicstyle=\footnotesize,caption={General top-down \emph{Mergesort}(A)},label={alg:Mergesort},captionpos=t,float,abovecaptionskip=-\medskipamount]
// Mergesort(A) sorts all elements of an array $A$
// according to a given comparison operator $\preceq$.
int $n, mid$

$n \gets |A|$ // number of elements in $A$
if $n > 1$
   $mid = \lfloor n / 2 \rfloor$ // split array into two halves
   Mergesort($A[0 \dots mid-1]$) // left recursion with $mid$ element
   Mergesort($A[mid \dots n-1]$) // right recursion with $n - mid$ elements
   $A[0 \dots n-1] \gets$ Merge($A, mid, n - mid$)
end if
\end{lstlisting}

\begin{lstlisting}[basicstyle=\footnotesize,caption={Traditional \emph{Merge}\_buffered($A, n_1, n_2$)},label={alg:Merge_buffered},captionpos=t,float,abovecaptionskip=-\medskipamount]
// Merge_buffered($A, n_1, n_2$) merges two sorted sequences
// into a single sorted sequence $A[0 \dots n_1+n_2-1]$.
// - First sorted sequence is in $A[0 \dots n_1-1]$.
// - Second sorted sequence is in $A[n_1 \dots n_1+n_2-1]$.
// It runs in $\mathcal{O}(n)$ time using $\mathcal{O}(n)$ additional space.
int $n$
typeof($A$) $R$

$n \gets n_1 + n_2$ // total number of elements
if $n > 1$
   $R \gets$ allocate_memory($n \times$typesize($A$))
   $\dots$ // sequentially merge both array parts into R
   $A[0 \dots n-1] \gets R[0 \dots n-1]$ // copy results back
   free_memory($R$)
end if
\end{lstlisting}

Before focusing on the actual \emph{in-place} variant of \emph{Merge}
and accordingly \emph{Mergesort}, we have to clarify when an algorithm can be
called \emph{in-place}. If it needs no additional memory but
also only a constant amount, it is called \emph{strictly
in-place}. But since such an $\mathcal{O}(1)$ requirement is often
too restrictive and would exclude any algorithm with a recursive nature,
which would already need an $\mathcal{O}(\log n)$ additional space for
the recursion stack itself, the following definition is generally
approved and can, e.g., be found in~\cite{mybook:Stepanov} (Def.~11.6 on page 215):

\begin{definition}\label{def:inplace}
An algorithm for an input length $n$ is called \emph{in-place} if
it uses $\mathcal{O}((\log n)^c)$ additional space (where $c$ is a constant).
\end{definition}

For the new \emph{in-place} \emph{Mergesort} the usual but
non-\emph{in-place} merging routine is replaced with the new
\emph{in-place} variant presented in Algorithm~\ref{alg:Merge_inplace}.
It utilizes both prerequisites
from Section~\ref{sec:prerequisites} in the following way:
The new \emph{in-place} merging takes an input array that is then
subdivided by an index $i$ into two already sorted subsequences.
These usually originate from the left and right \emph{Mergesort} after
these recursions have been completed. In principle, this splitting point
$i$ can be chosen arbitrarily, but for performance reasons, it should
be close to the middle of the input array. Using the \emph{co-ranking}
functionality from Section~\ref{sec:co-ranking}, we determine the
corresponding \emph{co-ranks} $j$ and $k$ for this $i$ in both
subsequences. As already mentioned previously, this can be done
in $\mathcal{O}(\log n)$ time for a total of $n$ elements using only
constant space. Once these
\emph{co-ranks} are known, the array is then logically partitioned
such that all array elements between these two indices
are considered to form one consecutive middle array. This middle array
is then rotated either by the resulting offset $k$ to the right or
alternatively by $i - j$ to the left\footnote{Since $i - j = k$ due to
the invariant of the \emph{co-ranking} functionality, it does not matter
if we rotate $k$ to the left or to the right. This might sound
counter-intuitive at first, but because rotating \textcolor{brown}{$r$}
to the left is equivalent to rotating
\textcolor{blue}{$n$}$-$\textcolor{brown}{$r$} to the right (see
note~\ref{note:rotate_leftright}), we get
\textcolor{blue}{$(i - j) + k$}$-$\textcolor{brown}{$k$} $= i - j =$
\textcolor{brown}{$k$}.}.
Utilizing the time- and space-optimal
array rotation from Section~\ref{sec:array_rotation},
this can be achieved \emph{in place}. Note that this array rotation
essentially moves all ``small'' elements to the left of $i$ and all ``large''
elements to its right. More precisely, afterward, all elements left of
$i$ precede $A[i]$, and all elements right of $i$ succeed $A[i-1]$.
In fact, this is guaranteed by the stable design of the
\emph{co-ranking} functionality (see~\cite{paper:MergeTheory} for more
details). Thus, the rotation does not need to care about stability anymore.
Finally, we can recursively call this \emph{in-place} merging
routine again on both array halves, which itself are already subdivided
into two arrays at $j$ and $k$ respectively. As usual, this recursion
stops whenever fewer than two elements are remaining.
Algorithm~\ref{alg:Merge_inplace} presents this new and recursive
\emph{in-place merge} approach.

\begin{lstlisting}[basicstyle=\footnotesize,caption={New \emph{Merge}\_inplace$(A,n_1,n_2)$},label={alg:Merge_inplace},captionpos=t,float,abovecaptionskip=-\medskipamount]
// Merge_inplace$(A,n_1,n_2)$ merges two given sorted sequences
// into a single sorted sequence in a memory-efficient way.
// It runs in $\mathcal{O}(n \log n)$ time using only $\mathcal{O}(n)$ comparisons.
// Space requirement is only $\mathcal{O}(\log n)$ due to recursion else $\mathcal{O}(1)$.
int $i, j, k$
typeof(A) $B, M$

if $(n_1 > 0)$ && $(n_2 > 0)$
   $B \gets \&A[n_1]$ // second array starts after the first one
   $|B| \gets n_2$
   // input 'i' is the rank in the nonexistent array 'C'
   $i \gets n_1$
   $(j, k) \gets$ Co_rank$(i, A, B)$
   // rotate middle array by offset $i - j = k$
   $M \gets \&A[j]$
   $|M| \gets (n_1 - j) + k$
   optimal_rotate$(M, k)$
   // recursively merge the resulting halves in place
   Merge_inplace$(A,j,n_1-j)$
   Merge_inplace$(B,k,n_2-k)$
end if
\end{lstlisting}

\begin{figure}
   \begin{center}
      \input{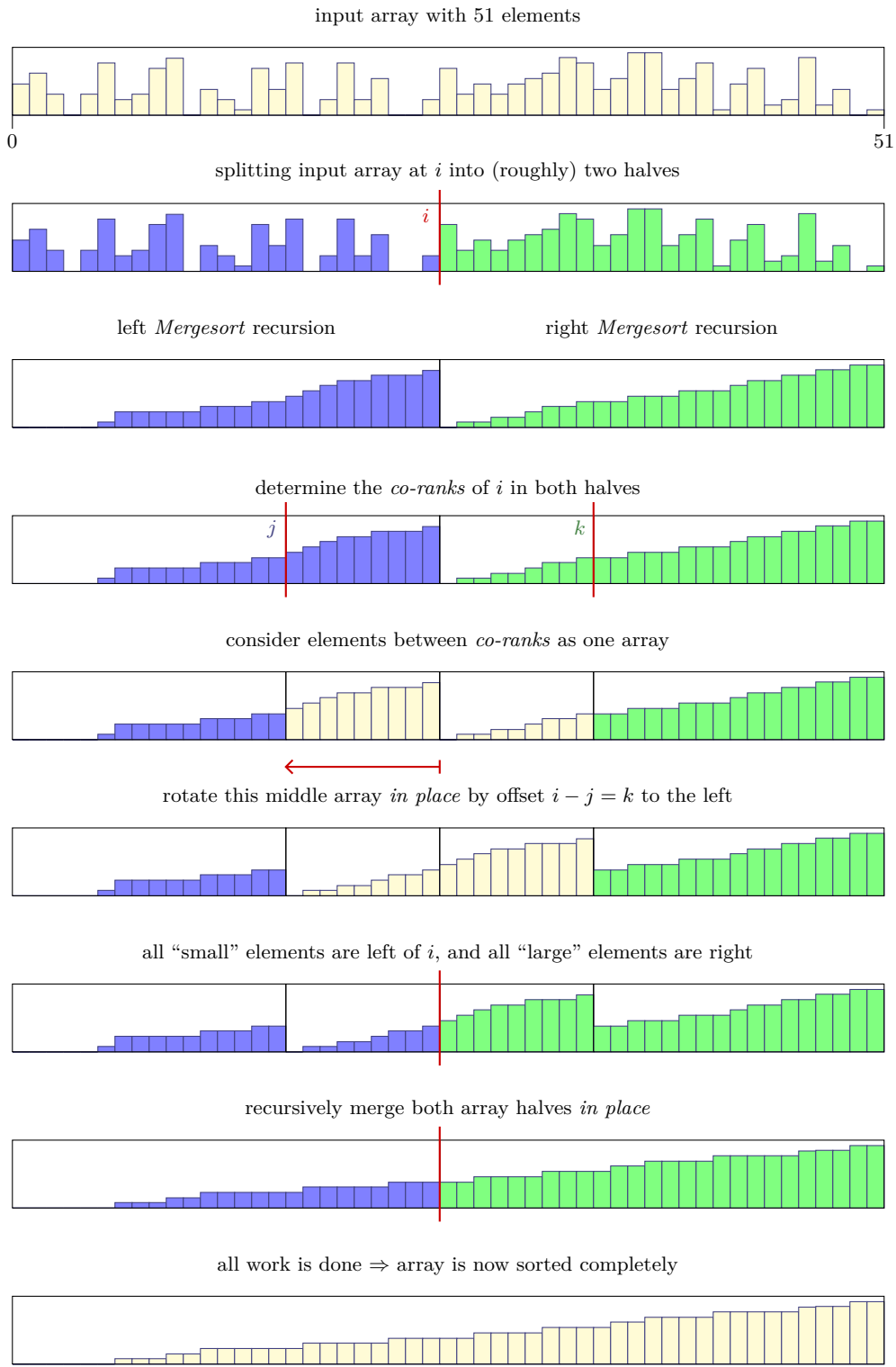}
   \end{center}
   \caption{Visualized example of the new \emph{in-place} \emph{Mergesort}}
   \label{fig:inplace_Mergesort}
\end{figure}

To facilitate better understanding, the entire \emph{in-place Mergesort}
scheme is also illustrated in Figure~\ref{fig:inplace_Mergesort}.

\subsection{Complexity analysis}
\label{sec:analysis}

Just like a normal \emph{Mergesort} routine, this new \emph{in-place} merging
requires $\mathcal{O}(\log n)$ recursion steps. Within each such recursion
step, it calls the \emph{co-ranking} functionality once. As explained in
Section~\ref{sec:co-ranking}, this \emph{co-ranking} executes
$\mathcal{O}(\log n)$ comparisons. Additionally, merging rotates the middle
array, which---despite requiring $\mathcal{O}(n)$ time to move
elements---does no element comparisons whatsoever. In total, this leads to
the following recursion formula for the number of comparisons in our
\emph{in-place} merging:

\begin{equation}\label{eqn:recurrence}
   \begin{array}{ll}
      T(0) & = 1 \\
      T(n) & = \mathcal{O}(\log n) + 2 \cdot T(\frac{n}{2}) \\
   \end{array}
\end{equation}

The popular \emph{master theorem}, which is typically used to solve
such recurrence relations, does not cover this specific
case. It can, however, be solved using the more general
\emph{Akra-Bazzi} method~\cite{paper:AkraBazzi}:
Matching our equation~(\ref{eqn:recurrence}) to their
$u_n = g(n) + \sum_{i=1}^k a_i u_{\lfloor \frac{n}{b_i} \rfloor}$ in (1)
yields $g(n)=\mathcal{O}(\log n)$,
$k = 2$, $a_i = 1$, $b_i = 2$ with $u_0 = 1$. Determining
the value of $p$ such that $\sum_{i=1}^k a_i b_i^{-p} = 1$ leads to
$p = 1$. Plugging this $p$-value into the \emph{Akra-Bazzi} equation
reveals the solution $T(n) \in \mathcal{O}(n)$.
Essentially, this means that this new \emph{in-place} merging routine
is still using only a linear number of comparisons, just like the usual
non-\emph{in-place} variant, and as such, it remains comparison-optimal.

\begin{note*}
   The current C++ standard already defines a
   \emph{std::inplace\_merge} function within its algorithm library.
   Concerning complexity, it states: ``\emph{Exactly N-1
   comparisons \dots if enough additional memory is available,
   $\mathcal{O}(N \log (N))$ comparisons otherwise.}''
   With our new \emph{in-place merge} algorithm,
   this can now always be improved to $\mathcal{O}(N)$ comparisons.
\end{note*}

This new \emph{in-place Merge} can be used directly for the general
top-down \emph{Mergesort} approach by replacing the \texttt{Merge}
call in Algorithm~\ref{alg:Mergesort} with \texttt{Merge\_inplace}.
The linear number of comparisons for the \emph{in-place Merge} in
conjunction with the usual $\mathcal{O}(\log n)$ recursion levels of
\emph{Mergesort} leads to the better-known recurrence
$T(n) = \mathcal{O}(n) + 2 \cdot T(\frac{n}{2})$,
which can be solved using the \emph{master theorem}, resulting in
$\mathcal{O}(n \log n)$. As such, the traditional, non-\emph{in-place}
\emph{Mergesort} and this new \emph{in-place Mergesort} asymptotically
utilize the same number of comparisons.

Regarding the running time, we also need to take into account the array
rotations, i.e., the movement of elements. Considering this, the new
\emph{in-place Merge} itself does a work of $\mathcal{O}(n \log n)$.
On that account, the total running time of
\emph{in-place Mergesort} consequently becomes $\mathcal{O}(n \log^2 n)$.

It should be noted that the idea of using rotations for an
\emph{in-place Merge} is not new. Some of the algorithms mentioned in
the related work make use of it. However, finding the rotation offset
was previously done utilizing only a linear search, leading to
$\mathcal{O}(n \log n)$ comparisons for the \emph{Merge} step and consequently
$\mathcal{O}(n \log^2 n)$ comparisons for the resulting \emph{Mergesort}.
Replacing such a linear search with the \emph{co-ranking} search
improves these numbers by a factor of $\mathcal{O}(\log n)$ comparisons,
eventually leading to our comparison-optimal solution.

Both prerequisites work with $\mathcal{O}(1)$ additional space and
the usual recursion necessitates only an additional $\mathcal{O}(\log n)$
space for the stack. For this reason, the new merging routine can be
classified as \emph{in-place}, albeit not \emph{strictly in-place},
according to Definition~\ref{def:inplace}.

\section{Performance evaluation}
\label{sec:results}

We have implemented the presented \emph{in-place Mergesort} algorithm
in plain ANSI-C. Our implementation supports arbitrary comparators using
a generic comparison function, just like the standardized \emph{qsort()}
function. This way, we can not only compare our implementation to an
optimized baseline but also simply
add a counter increase to our comparison function, which makes it
easy to count the total number of comparisons needed to sort a given
array. We then fit this counted number of comparisons to the formula
$c \cdot n \cdot \log_2 n$ to determine the practical factor
$c$ for different array sizes. The traditional, buffered \emph{Mergesort}
achieves a factor of $0.874 \leq c \leq 0.958$, and for the novel,
\emph{in-place Mergesort} this reveals a factor of $2.215 \leq c \leq 2.523$.
Deducing from the resulting ratio, our novel \emph{Mergesort}
executes $2.5$ times more comparisons than the traditional variant,
which is certainly a constant. In addition to the already theoretically
established number of comparisons of $\mathcal{O}(n \log n)$
(see Section~\ref{sec:analysis}), these experimental results provide
further empirical evidence for this optimal scaling behavior.

\begin{figure}
   \begin{center}
      \input{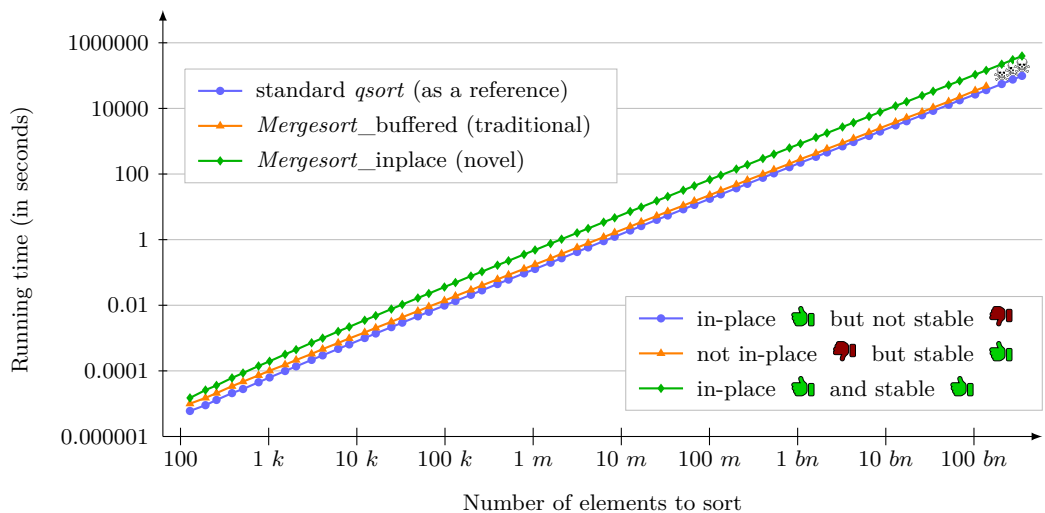}
   \end{center}
   \caption{Performance comparison of different sorting algorithms}
   \label{fig:sorting_performance}
\end{figure}

Furthermore, we measured the resulting performance of our implementation.
We used numbers in double-precision floating-point format with a size
of 64 bits for the array to be sorted. The values of these numbers were
generated by the \emph{drand48()} function from the standardized
pseudo-random number generator. This generator uses an internal state
of 48 bits and 48-bit arithmetic for its calculations, yielding a high
precision of the generated values as well as a large period.
The \emph{GCC} compiler, version 13.2.0, was used to build the measurement
and test application, using only the general optimization option -O3.
The measurements were conducted on the HPC system 
``Hilbert'' at the Heinrich Heine University D{\"u}sseldorf in Germany,
on a large
quad-socket Intel Xeon Gold 6150 ``Skylake'' system equipped with 3~TiB
of shared memory. After each measurement, the results were successfully
checked for correctness.

Figure~\ref{fig:sorting_performance} shows actual performance results
for various numbers of elements (both power-of-two and non-power-of-two)
of three different sorting implementations:
(1) standard \emph{qsort()} as an optimized reference;
(2) \emph{Mergesort}\_buffered as implemented traditionally;
and (3) \emph{Mergesort}\_inplace as presented in this work.

The first one is the
\emph{qsort()} implementation of the standard C library, in this case
the GNU C library version 2.17 using \emph{Quicksort}. This serves
as a highly optimized baseline for reference.
While \emph{qsort()} is usually fast, its
timings vary quite a lot due to the problem of choosing an appropriate
pivot and can be even orders of magnitude slower than alternatives.
In order to still show somewhat meaningful measurement results, we
randomized the input
data, i.e., a new seed value for every measurement, conducted several
measurements and selected the median runtime. In contrast, the other two
implementations do not have this problem, as they show repeatable
performance characteristics with only minor runtime variations.
More precisely, repeating the same measurement setup hundreds of times
with our new \emph{in-place Mergesort} implementation yields reproducible
timing results with a variance of only $\pm 0.2\%$.

The second candidate shown in this performance chart is a traditional
\emph{Mergesort} implementation that uses a second buffer needed for
the \emph{Merge} step, as shown in Algorithm~\ref{alg:Merge_buffered}.
The third implementation represents our new \emph{in-place Mergesort}, as
depicted in Algorithm~\ref{alg:Merge_inplace}, which---besides logarithmic
stack space for the recursions---does not need additional memory.

The \emph{qsort()} reference implementation generally uses the
\emph{Quicksort} algorithm, originally developed by Tony Hoare in 1959.
However, it is highly optimized and uses \emph{insertion sort},
i.e., another sorting algorithm, for a small number of elements below
a certain threshold. Besides that, it implements a better pivot selection
strategy and other optimizations, such as all-inlining, a non-recursive
stack, cache-friendly data layouts and streamlined data accesses.
So while we can expect better performance, it is important to remember
that \emph{Quicksort} implementations and, hence, also \emph{qsort()} are
not stable. As such,
the C standard restricts its use and states that if two keys compare as
equal, their order in the sorted array is undefined.

On the other hand, both shown \emph{Mergesort} implementations are stable
and straightforward implementations without any special optimizations.
Due to the threshold-selected algorithm choice in the reference
implementation, the buffered \emph{Mergesort} is around 50 to 100\%
slower than \emph{qsort()} for small numbers of elements
and only around 26 to 40\% for large numbers of elements.
This is quite good, considering that many of the optimizations applied
to \emph{qsort()} can also be applied to both \emph{Mergesort}
implementations. For example, we can make the new \emph{Merge} algorithm
\emph{memory-adaptive} by setting some threshold, i.e., the amount
of additional memory that we are willing to spend in exchange for
some performance gains, and switching to a different sorting algorithm
whenever it encounters a smaller array, e.g., during recursion.
Nevertheless, any such optimizations regarding an actual implementation are
outside the scope of this paper, and reintroducing additional memory
for improved speed would essentially revoke the quality characteristic
of working \emph{in place}.

Ignoring any optimizations, a fairer comparison of the traditional
buffered \emph{Mergesort} and the new \emph{in-place} variant should
reveal the actual performance overhead needed in exchange for
reducing the additional memory footprint from $\mathcal{O}(n)$ down
to $\mathcal{O}(\log n)$. While there is a factor of
$\mathcal{O}(\log n)$ for the theoretical performance difference,
the new \emph{in-place Mergesort} actually becomes only a factor between
2.0 and 3.1 slower than the traditional \emph{Mergesort} implementation,
or a factor of 2.6 to 4.0 slower than the optimized \emph{Quicksort}.
Although this is expected theoretically for the number of comparisons
executed (see Section~\ref{sec:analysis}), this is a rather positive
surprise for these practical measurements due to the still necessary
array rotations (see Section~\ref{sec:array_rotation}). A straightforward
analysis with the statistical profiler \emph{GNU gprof} determined that,
e.g., in a sorting run with 62 million elements, 10.9 seconds of the
\emph{in-place merge} were spent in the co-ranking, while 13.3 seconds
were spent in the array rotation.
Increasing this problem size by a factor of 10 increases these numbers
to 121.6 seconds ($\times$ 11.2) for the co-ranking and 176.7 seconds
($\times$ 13.3) for the array rotation. A curve fitting revealed a
constant factor of $6.1$ ($\pm 0.5$) for the $\mathcal{O}(n \log n)$
co-ranking part and a constant factor of $0.25$ ($\pm 0.05$) for the
$\mathcal{O}(n \log^2 n)$ array rotation part.
While the actual work due to the array rotations is a factor
of $\mathcal{O}(\log n)$ worse than for the number of comparisons, the
practical constant is much smaller---the actual timings are almost on par
(43\% vs. 57\%). This shows that the array rotations are executed
efficiently, as they are mostly cache-friendly, and thus impact our
\emph{in-place} implementation less than the relatively expensive
comparisons.

Besides randomized input data, we also measured our \emph{in-place Mergesort}
implementation with specially generated distributions for the input data,
such as pre-sorted or
alternating elements simulating favorable and unfavorable inputs.
Regardless of the tried distributions, it still gives very
repeatable timing results with differences of at most $5.0\%$.
Just for comparison, WikiSort timings vary considerably by more than $100\%$.
Thus, this alleged competitor is very sensitive to the actual data content
and can even fail if its impractical restrictions (enough
additional memory available and the presence of many duplicate data values)
are not fulfilled. For this reason, we could not conduct a fair
performance comparison. Contrary to WikiSort, our proposed algorithm is
insensitive to the actual data, truly \emph{in place}, generally
applicable and always works without any restrictions.

Last but not least, we observe that the traditional \emph{Mergesort}
fails at 206 billion elements and beyond, marked with
\includegraphics[width=3mm]{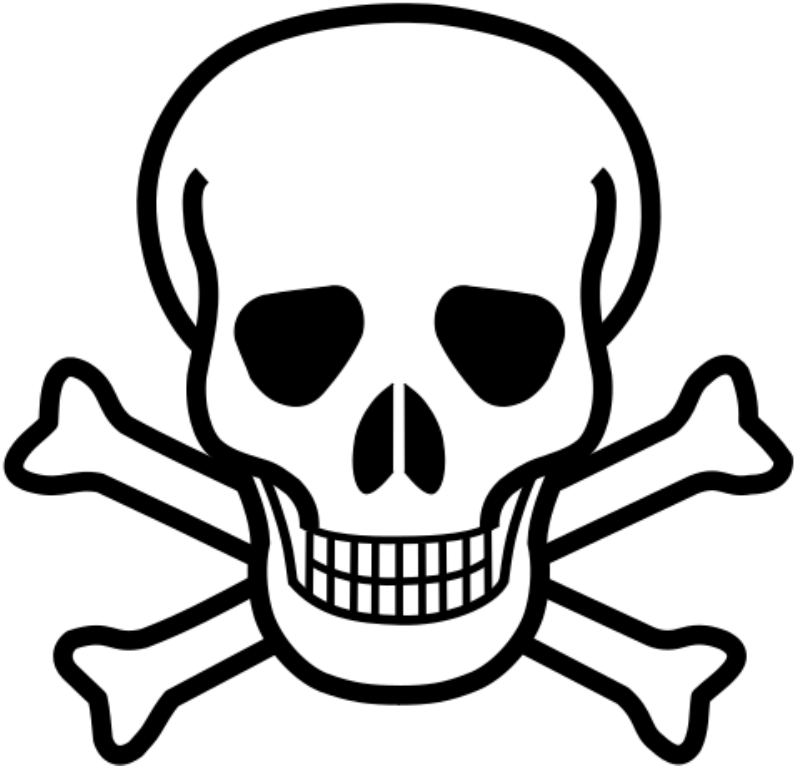},
as this would require the
allocation of more than 1.5~TiB of additional memory, which even
this large shared memory system simply cannot provide. The novel
\emph{in-place Mergesort}, on the other hand, has no such problems
and can successfully deal even with 350 billion elements, which
corresponds to an input data set of 2.545~TiB or almost all available
memory. In general, our
\emph{in-place Mergesort} enables the stable sorting of huge data
sets well beyond half the available main memory capacity.
Regardless of performance, none of the competitors from
Table~\ref{table:sortprop}, i.e., \emph{Quicksort}, traditional
\emph{Mergesort}, \emph{Heapsort}, can achieve this accomplishment.

\section{Conclusion}
\label{sec:conclusion}

In this paper, we present a novel approach to create a \emph{Mergesort}
algorithm that works \emph{in place}, requiring only $\mathcal{O}(\log n)$
additional space. This new approach utilizes just two existing prerequisites,
which makes it easy to comprehend and implement. While often lacking by
competitors,
this new \emph{in-place Mergesort} is stable by design. Although the
theoretical work scales with $\mathcal{O}(n \log^2 n)$, the number of
comparisons achieves the optimal lower bound of $\mathcal{O}(n \log n)$.
Furthermore, we implemented this new algorithm and measured its practical
performance on one of the biggest shared memory systems currently in service.
The evaluation results revealed that its actual running time is merely a
small factor slower than both a traditional buffered \emph{Mergesort}
implementation and a highly optimized \emph{qsort()} implementation. These
first promising results were achieved without any optimizations worth
mentioning. This leaves further room for improvements, such as making the
algorithm \emph{memory-adaptive}, to reduce this small overhead even
further. The runtime of our implementation is insensitive to the input
data and gives reproducible timing results. Moreover, it does not depend
on any restrictions, such as the availability of many duplicate values in
the input data, and is therefore generally applicable.
Altogether, it is now finally possible to
have a sorting algorithm that is not only simple to implement with small
factors in practice but also memory-efficient with just
$\mathcal{O}(\log n)$ additional space, stable and yet always
comparison-optimal, i.e., $\mathcal{O}(n \log n)$.

\section*{Acknowledgment}

Computational infrastructure and support were provided by
the Centre for Information and Media Technology at Heinrich Heine
University D\"usseldorf.

\bibliography{references}

\end{document}